\documentstyle[preprint,aps]{revtex}

\begin{document}

\title{ Dimensional versus cut-off renormalization  and the
nucleon-nucleon interaction\thanks{To appear in Physical Review C}}

\author {Angsula Ghosh$^1$,  Sadhan K. Adhikari$^1$
   and B. Talukdar$^{1, 2}$}
   \address{ 
  $^1$Instituto de F\'\i sica Te\'orica, 
Universidade Estadual Paulista, \\ 01405-900 S\~{a}o Paulo,  S\~{a}o Paulo, 
Brasil \\
$^2$Department of Physics,  Visva Bharati,  Santiniketan 731235,  India}

\date{\today }

\maketitle

\begin{abstract}

The  role of dimensional regularization  is discussed and compared with that
of cut-off regularization in some quantum mechanical problems with
ultraviolet divergence in two and three dimensions with special emphasis on
the nucleon-nucleon interaction.  Both types of renormalizations are
performed for attractive divergent  one- and two-term separable potentials,
a divergent tensor potential, and   the sum of  a delta function and its
derivatives.   
 We allow energy-dependent couplings,
and determine the form that these couplings should take if equivalence 
between the two regularization schemes is to be enforced.
We also perform  renormalization of an attractive separable
potential superposed on an analytic divergent  potential.

{\bf PACS Numbers 21.30.-x,  03.65.Nk,  11.10.Gh}

\end{abstract}

\newpage

\section{Introduction\label{s1}}

Ultraviolet divergences appear in exact as well as perturbative treatments of
the nonrelativistic quantum mechanical two-body problem in momentum space
interacting via two-body potentials with certain singular behavior at short
distances \cite{ad,adx,jac,lat,coh1,coh2,coh3,van,lep,bir,eft2,eft3,wei} 
in two and three
space dimensions.  Similar divergences appear in perturbative quantum field
theory and are usually treated by renormalization techniques \cite{wil,dim}.
There are several variants of renormalization which employ different types of
regularizations,  such as,  the cut-off, and dimensional regularizations.
Unless there is some symmetry violation in performing regularization,  in
perturbative field theory,  both regularization schemes are expected to lead
to the same renormalized result at low energies. The closely related
technique   of discretization on the lattice in such field theoretic problems
also should lead to equivalent results. Exactly as in quantum field theory,
the ultraviolet divergences in quantum mechanics can be treated by
renormalization. Three schemes have been used for the purpose: cut-off
regularization \cite{ad,adx,jac,coh1,coh2,coh3,bir,eft2},  
dimensional regularization \cite{coh2,lep,eft2},
and discretization on the lattice \cite{lat}.  For the simplest
$\delta$-function potential all three approaches lead to identical result.

Recently,  cut-off \cite{ad,coh1,coh2,coh3,bir,eft2} 
and dimensional \cite{coh2,bir,eft2}
regularizations  have been successfully used by several workers  in quantum
mechanical problems in $S$ and higher partial waves in the context of
nucleon-nucleon interaction.  Although,  both regularization schemes have
been successfully used for the purpose of renormalization in quantum
mechanical problems with ultraviolet divergence, their equivalence can not be
demonstrated except in the simplest problems. In this work   we consider 
several problems with ultraviolet divergence, allow energy-dependent 
bare couplings, and find the forms that these couplings must take in order to 
obtain equivalence between the two regularization schemes.

We consider four  potentials for illustration. The simplest is the minimal
potential in a general partial wave considered recently \cite{ad}. 
In momentum space
this potential possesses only the threshold behavior and is given by
the following one-term separable form
\begin{equation}
V(p, q)=p^L\lambda_L q^L,\label{1}
\end{equation} 
with $L$ the angular momentum. For $L=0$,  this potential is the usual
$\delta$-function  potential.  As the  Lippmann-Schwinger
equation has the same generic form in all partial waves,  the ultraviolet
divergence of this potential becomes stronger and stronger as $L$ increases.
 Next we consider a
two-term separable potential of the form
\begin{equation}
V(p, q)= \lambda_0 u(p) u(q)+\lambda_2 v(p) v(q),
\label{2a}\end{equation}
 where $u(p)$ and $v(p)$ are each divergent form factors of the type $p^L$, 
considered in the first potential. We also consider the potential
\cite{coh1,coh2,bir,eft2} 
\begin{equation}\label{3}
V(p, q) = \lambda_1 + \lambda_2 (p^2+ q^2).
\end{equation} Potential (\ref{3}) is the sum of a $\delta$ function and its
second derivatives. This potential  is interesting as it appears as a piece
in the low-energy nucleon-nucleon potential derived from effective field
theory and has received attention recently \cite{eft2,wei}. 
Finally,  we consider the following minimal tensor nucleon-nucleon potential 
possessing ultraviolet divergence
\begin{equation}\label{40}
|V_{LL'}(p, q)| \equiv
\left(\begin{array}{cc} V_{00}(p, q)   &   V_{02}(p, q)\\
V_{20}(p, q)  & V_{22}(p, q)  \\ \end{array}\right)=
 \left(\begin{array}{cc} \lambda_0   &   \lambda_1 q^2\\
\lambda_1 p^2  & \lambda_2 p^2q^2  \\ \end{array}\right),
\end{equation}
where $\lambda_0$  and $\lambda_2p^2q^2$ are  divergent 
$S$ and $D$ wave parts of the nucleon-nucleon potential  (\ref{1}).
In Eq. (\ref{40}) 
the angular momentum labels $L$ and $L'$ are explicitly shown.
The term involving $\lambda_1$ provides
 the $S-D$ coupling. When $\lambda_1=0$, 
the $S$ and $D$ waves decouple and we essentially have potential (\ref{1}).
The functions $p^2$ and $q^2$ are the threshold factors for $L=2$.

The nucleon-nucleon potential derived from a chiral Lagrangian formulation 
of effective field
theory contains usual finite-range potentials superposed on divergent
potentials containing delta-function and derivatives (gradients)
and can be written as 
\cite{coh1,coh2,bir,eft2} 
\begin{equation}\label{div}
V(p, q)= V_f(p, q)+
\lambda_1+\lambda_2(p^2+q^2)+\lambda_3 p^2 q^2+\lambda_4(p^4+q^4)
+ \ldots,
\end{equation}
where $V_f(p, q)$ represents usual finite-range parts of the potential. 
The configuration-space derivatives of the $\delta$ function appear as 
powers of momenta in momentum space.
Potential (\ref{3}) is just a part  of Eq. (\ref{div}).
Though one can renormalize the divergent parts separately,  it is not clear
that a potential,  such as (\ref{div}), 
 containing a divergent and a finite-range part can be successfully
renormalized.
Though a general answer to this question involving local finite-range
potentials 
may involve a complicated numerical 
calculational scheme,  we would like to address this point in a much simpler
context,  where we take the finite-range part to be an attractive
separable potential frequently used to simulate the nucleon-nucleon
interaction.  We perform an  analysis to show that when a  divergent
potential  of the form $\lambda$ or $\lambda p^2 q^2$ is summed to an 
attractive
separable potential,  the renormalized scattering $K$ or $t$ matrix 
leads to physically plausible results for the nucleon-nucleon system at low
energies.

The plan of our work is as follows. We perform renormalization of potentials
(\ref{1}),  (\ref{2a}),  (\ref{3}) and (\ref{40}),  in Secs.
\ref{s2a},  \ref{s2b},  \ref{s2c} and \ref{s2d},  respectively,  using both
dimensional and cut-off regularizations.  In Sec. \ref{s2a} we consider the
scattering problem in both two and three dimensions.  The  potentials
(\ref{2a}),  (\ref{3}) and (\ref{40}) are  mostly of interest in nuclear
physics and hence we shall be limited to only the three-dimensional case in
Secs. \ref{s2b},  \ref{s2c} and \ref{s2d}.  In Sec. III we present
results for the renormalization of a  divergent potential  of the form
$\lambda$ or $\lambda p^2 q^2$ added  to an attractive separable potential.
Finally,  in Sec. IV a brief summary  of the present work is presented.

\section{Regularization and Renormalization}

The partial-wave Lippmann-Schwinger equation for the $K$ matrix
$K_L(p, q, k^2)$,  at center of mass  energy $k^2$,  is given,  in 
dimension $d$,  by
\begin{eqnarray}
K_L(p, k, k^2) = V_L(p, k)+ {\cal P}
\int q^{d-1} dq V_L(p, q)
G(q; k^2)K_L(q, k, k^2),
\label{2} \end{eqnarray} 
with the free Green function $G(q; k^2)=(k^2-q^2)^{-1}, $ in units
$\hbar=2m=1$,  where $m$ is the reduced mass;  ${\cal P}$ in  Eq. (\ref{2})
denotes principal value prescription for the 
integral and the momentum-space integration limits are from 0 to $\infty$. 
The (on-shell) scattering amplitude 
$t_{L}(k)$ is defined  by 
\begin{equation}\label{33}
\frac{1}{t_L(k)}=\frac{1}{K_L(k^2)}+i\frac{\pi}{2}k^{d-2},
\end{equation}
where $K_L(k^2)\equiv K_L(k, k, k^2)= -(2/\pi)(\tan \delta_L/k)$ with 
$\delta_L$ the phase shift.
All scattering observables can be calculated using ${t_L({k})}.$
Though we are considering a general 
dimension $d$,  we shall  be limited in the present work to $d=2$ and 3.
Most of our results can be generalized to higher dimensions. 
The
condition of unitarity is given by
\begin{equation}
\Im t_L(k) = -\frac{\pi}{2}k^{d-2} |t_L(k)|^2,\label{uni}
\end{equation}
where  $\Im$ denotes the imaginary part. Here  we employ a $K$-matrix
description of scattering. Then the renormalization algebra will involve only
real quantities and we do not have to worry about unitarity which can be
imposed later via Eq. (\ref{33}). This is the simplest procedure to follow,  as
all  renormalization schemes preserve unitarity.

\subsection{ \label{s2a}
The Minimal Potential}

Minimal potential (\ref{1}) has been renormalized by cut-off regularization 
 in three dimensions in  \cite{ad} using the $t$ matrix approach.
  Here we present a brief account of that
work with appropriate generalization to the $K$ matrix approach in both 
two and three dimensions.
We also perform dimensional regularization with this potential 
 and discuss its consistency with 
cut-off regularization. 
For  minimal potential (\ref{1}),  the $K$ matrix of  Eq. (\ref{2})
permits the following analytic solution
\begin{equation}
K_L(p', p, k^2)=p'^L \tau_L(k) p^L,
\label{4}
\end{equation}
with the $\tau$ function defined by
\begin{eqnarray}
\tau_L(k) & = &  [\lambda_L^{-1}-I_L(k)]^{-1} \label{4a}\\
I_L(k)& = & {\cal P}  \int q^{d-1} dq  q^{2L} G(q; k^2).\label{4b}
\end{eqnarray} 
Integral $I_L(k)$ of  Eq. (\ref{4b}) possesses ultraviolet divergence. 
As the $\tau$ function completely determines the $K$ matrix and as the 
divergent terms are contained in it,    we 
consider  the renormalization of this  function.  

Because of the ultraviolet divergence in Eq. (\ref{4b}), 
 some regularization is needed to give meaning to it. 
We  use the following regularized Green
function with a sharp cut-off
\begin{eqnarray}
G_R(q, \Lambda; k^2) = (k^2-q^2)^{-1} \Theta(\Lambda-q), \label{9a}
\end{eqnarray} with
$\Theta(x)=0$ for $x < 0$ and =1 for $x > 0$.  In   Eq. (\ref{9a}),   $\Lambda
(>>k)$ is a large but finite quantity.  
In the end,   the limit $\Lambda
\to \infty$ has to be taken.  Finite results for physical magnitudes,  as
$\Lambda \to \infty$,  are obtained only if the coupling $\lambda_L$ is 
understood to be
 the so called bare coupling $\lambda_L(k, \Lambda)$.

In order to proceed with the choice of the bare coupling we have to consider
a specific value of dimension $d$. First,  we consider $d=3$.  The choice of
the bare coupling can be found by inspection of the following regularized
form of Eq. (\ref{4b})
\begin{eqnarray}
I_{RL}(k, \Lambda) & \equiv & {\cal P} \int q^2 dq q^{2L}
G_R(q, \Lambda; k^2)\\ & = &
-\left[  \sum_{j=0}^L  \frac{k^{2(L-j)}\Lambda^{2j+1}}{2j+1}
+\frac{k^{2L+1}}{2}\ln\left|\frac{\Lambda-k}{
\Lambda+k}\right|
\right].\label{r3}
\end{eqnarray}
In the   large $\Lambda$ $(>> k)$ limit,  the logarithmic term in
Eq. (\ref{r3}) tends to zero and  
\begin{eqnarray}
\lim_{\Lambda \to \infty}I_{RL}(k, \Lambda)=-
 \sum_{j=0}^L  \frac{k^{2(L-j)}\Lambda^{2j+1}}{2j+1}.\label{101}
\end{eqnarray}
All  terms in   Eq. (\ref{101}) diverge as $\Lambda \to
\infty.$ Except for $L=0$,  these divergent terms are momentum $(k)$
dependent.

For obtaining a finite renormalized $\tau$ function,  the coupling
$\lambda_L$ should be understood to be  the so called bare coupling
$\lambda_{L}(k, \Lambda)$. In Ref.
\cite{ad}  the following energy-dependent bare coupling was used:
\begin{eqnarray}
\lambda_{L}^{-1}(k, \Lambda) & = & -  \sum_{i=0}^L
\frac{k^{2(L-i)}\Lambda^{2i+1}}{2i+1} +
 \Lambda_{0L}(k^2), \label{x}
\end{eqnarray}
where  function $\Lambda_{0L}(k^2)$   can be used to introduce 
the physical scale(s) of the
system and characterizes the strength of the interaction. The quantity 
$1/\Lambda_{0L}(k^2)$ could be termed the renormalized strength or coupling
of the interaction.   We are taking 
the bare coupling to be energy dependent. 
In the present non-relativistic context this is not of
concern,  as the bare coupling is not an observable. We shall see that the 
renormalized $K$ matrix obtained after renormalization from this bare
coupling has the desired analytic properties in $k^2$. We shall see that 
this energy dependence is necessary, in order to obtain equivalent results
from cut-off and dimensional regularizations.

However, there are reasons to argue against the use of
energy-dependent bare couplings. In field theory they correspond to counter
terms which violate time reversal invariance,  and also destroy the usual 
Hermitean structure of the quantum-mechanical potential generating the $t$
matrix. In spite of these, we shall use energy-dependent bare couplings in
this study, as that seems to be a mean  for  obtaining equivalent results
from cut-off and dimensional renormalization
as the ultraviolet divergence is energy dependent in this case. 
The renormalized physical
result should be independent of the regularization schemes employed.
For $L> 0$, because of 
the additional freedom provided by the energy-dependent couplings,
we can obtain equivalent results from the two different regularization 
schemes.

Employing bare coupling (\ref{x}),
the regularized $\tau$ function of  Eq. (\ref{4a}) can now be rewritten as
\begin{eqnarray}
\tau_{RL}(k, \Lambda)& = &[\lambda_L^{-1}(k, \Lambda)-I_{RL}(k, 
\Lambda)]^{-1}, 
\label{8} \end{eqnarray}
where  for a finite $\Lambda$,  $I_{RL}(k, \Lambda)$ is a convergent
integral.  As $\Lambda \to
\infty$,   $\lambda_L^{-1}(k, \Lambda)$ of   Eq. (\ref{x}) has
the appropriate divergent behavior,   that cancels the divergent part
Eq. (\ref{101})  of
$I_{RL}(k, \Lambda)$.   As $\Lambda \to \infty$, 
one obtains the following renormalized $\tau$ function from  Eqs. (\ref{101}), 
(\ref{x}) and (\ref{8}):
\begin{eqnarray}
\tau_{{\cal R}L}(k) =  \lim_{\Lambda \to \infty}
[\lambda_L^{-1}(k, \Lambda)-I_{RL}(k, \Lambda)]^{-1} 
= \frac{1}{\Lambda_{0L}(k^2)}. \label{ex}
\end{eqnarray}
 In three dimensions, 
the low-energy scattering is usually parametrized by a few parameters as in
the effective-range expansion. Hence it is natural to take 
\begin{equation}
\Lambda_{0L}(k^2)= -1/a_L^{2L+1}+b_L^{1-2L}k^2+\ldots,\label{poly}
\end{equation}
where $a_L$ is the scattering length and $b_L$ is a range parameter.
These parameters are usually called physical scales as they measure the
physical observables,  such as,  a cross section.

Next,  we consider  $d=2$. 
Here the regularized integral $I_{L}(k)$, in the large $\Lambda$ limit
($\Lambda> > k$),
becomes \cite{coh2}
\begin{eqnarray}
I_{RL}(k, \Lambda) & \equiv & {\cal P} \int q dq q^{2L}
G_R(q, \Lambda; k^2)\\ & = &
-\biggr[-\ln \frac{k}{\Lambda}\biggr],  L=0 \\
  & =& -\biggr[\frac{\Lambda^2}{2}-k^2 \ln \frac{k}{\Lambda}
 \biggr],  L = 1\\ 
 & =& -\biggr[\frac{\Lambda
^4}{4}+\frac{k^2\Lambda^2}{2} -k^4 \ln \frac{k}{\Lambda}\biggr],  L=2
\end{eqnarray}
so that for a general $L$ we have 
\begin{eqnarray}
I_{RL}(k, \Lambda)=-
 \sum_{j=1}^L  \frac{k^{2(L-j)}\Lambda^{2j}}{2j}
+k^{2L}\ln \frac{k}{\Lambda}.\label{102}
\end{eqnarray}
In order to obtain a finite renormalized $\tau$ function,   the coupling
$\lambda_L$ should be understood to be 
 the so called bare coupling  defined,  for
example,   by
\begin{eqnarray}
\lambda_{L}^{-1}(k, \Lambda) & = & -  
 \sum_{j=1}^L  \frac{k^{2(L-j)}\Lambda^{2j}}{2j}
+k^{2L}\ln \frac{\Lambda_{0L}(k^2)}{\Lambda}, \label{x1}
\end{eqnarray}
where the function $\Lambda_{0L}(k^2)$  
can again be used to introduce  the physical scale(s) of the
system and characterizes the strength of the interaction as in the case 
with $d=3$.  If we use Eqs. 
 (\ref{102})  and (\ref{x1})  in  Eq. (\ref{8}) the following
renormalized $\tau$ function  is obtained in the limit $\Lambda \to \infty$:
\begin{eqnarray}\label{17a}
\tau_{{\cal R}L}(k) =-  
 \frac{1}{ k^{2L}\ln[k/\Lambda_{0L}(k^2)]}. \label{ex1}
\end{eqnarray}
The $\ln(k)$ dependence in Eq. (\ref{ex1}) is the proper low-energy 
momentum dependence in two dimensions \cite{adz}.
Expressions (\ref{ex}) and (\ref{ex1})  are the renormalized 
$\tau$ functions
obtained with cut-off regularization for $d=3$ and 2,  respectively.
Now it is realized that the use of energy-dependent bare coupling (\ref{x})
is essential 
for removing the ultraviolet divergences in the $\Lambda
\to \infty$ limit by cut-off regularization.

The above-mentioned problem can also be tackled with the help of dimensional
regularization \cite{dim}. 
In this procedure integral (\ref{4b}) is evaluated to 
yield \cite{g}
\begin{eqnarray} \label{30}
I_{RL}(k, d) & \equiv & {\cal P}\int q^{d-1}dq q^{2L}(k^2-q^2)^{-1}\\
&=& -\frac{1}{2}\Gamma\left(\frac{2L+d}{2}\right)
\Gamma \left(  \frac{2-2L-d}{2} \right)\Re\left[
{(-k^2)^{(2L+d-2)/2}}\right],  
\label{31}
\end{eqnarray}
where $\Re$ denotes the real part.  Integral  (\ref{30}) is divergent for
$d=2$ and 3 and  (\ref{31})  is the finite result valid for $0< (d+2L)< 2$. In
dimensional regularization,  Eq. (\ref{31}) is interpreted  to be an
extrapolation of the convergent result for small $(d+2L)$ $(< 2)$ to the
actual values of $(d+2L)$ $(\ge 2)$ for which the result is divergent. In odd
dimensions the dimensionally regularized  integral (\ref{31}) is zero because
it contains the real part of an imaginary quantity.  However,  for
dimensional regularization in even dimensions,  Eq. (\ref{31}) is used to
extract the divergent  part of $I_{RL}(k)$ as in Eq. (\ref{101}) and then
renormalization can be performed.

For $d=3$,  the dimensionally regularized
result (\ref{31}) is already finite:
\begin{eqnarray} \label{32}
I_{RL}(k) 
= 0,
\end{eqnarray}
 and in this case one does not need to introduce a 
 new energy-dependent bare coupling and from 
Eqs.  (\ref{4})  and (\ref{4a})
 one immediately obtains
\begin{eqnarray} \label{34}
\tau_{RL}(k)=\lambda_L,  \hskip 1cm \makebox{and} \hskip 1cm
K_{RL}(k^2)=\lambda_L k^{2L}. 
\end{eqnarray}
However, 
equivalence between the cut-off renormalized result (\ref{ex})
and the dimensionally
regularized result (\ref{34}) is obtained if the following energy-dependent 
bare coupling is used instead
\begin{equation}\label{bc}
\lambda_L(k) =
 1/\Lambda_{0L}(k^2).\end{equation}

For $d=2$,  the dimensionally regularized result still
contains infinities and a subtraction of these infinities 
is necessary before
obtaining a finite renormalized result. In this case,  as $d\to 2$,  
Eq.  (\ref{31})  can be rewritten as \cite{boi}
\begin{eqnarray}\lim_{\epsilon \to 0}
I_{RL}(k, \epsilon)= -k^{2L}\left[\frac{\Gamma(\epsilon)}{2}\right]
\Re[ (-k^2)^{-\epsilon}]
=-k^{2L}\left[
 \frac{1}{2\epsilon}
-\ln {k} -\frac{\gamma}{2}\right],\label{103}
\end{eqnarray}
where $\gamma=0.577..$ is the Euler number and 
$\epsilon = (1-d/2)$. In writing  Eq. (\ref{103}),  use has been made of the 
well-known limits
\begin{eqnarray}
\lim_{\epsilon \to 0} {\Gamma(\epsilon)}
 &\to & \frac{1}{\epsilon}-\gamma+{\cal O}(\epsilon)+\ldots\\
 \lim_{\epsilon \to 0} \Re[ (-k^2)^{-\epsilon}] & \to &
 1-2\epsilon \ln k + {\cal O} (\epsilon^2)+\ldots
\end{eqnarray}
and $\Gamma(1+x)=x\Gamma(x)$. In the limit $d\to 2$,  $\epsilon \to 0$ and 
 (\ref{103}) is divergent.
If we compare Eq.
(\ref{103})  with  Eq. (\ref{102})
we find that the $1/\epsilon$ pole in the dimensionally regularized integral
corresponds to the different divergences including a logarithmic divergence
 in the cut-off regularized integral. Equation (\ref{103})  contains the 
logarithm of $k$ $-$ a dimensional variable. The scale of the logarithm is
hidden in the $1/\epsilon$ term and appears when the divergence is canceled
with an appropriate choice of  bare coupling \cite{boi}. 
For performing renormalization with the dimensionally
regularized result (\ref{103}),  one should choose the bare coupling as 
\begin{equation}\label{eq2}
\lambda_L^{-1}(k, \epsilon)=-
 \frac{k^{2L}}{2}\left[\frac{1}{\epsilon}-\gamma\right]
+k^{2L}\ln {\Lambda_{0L}(k^2)}.
\end{equation}
In the limit $\epsilon \to 0 $,  using  Eqs. (\ref{103})  and (\ref{eq2})
in
\begin{equation}
\tau_{RL}(k)=\lim_{\epsilon \to 0}[\lambda_L^{-1}(k, \epsilon)
- I_{2L}(k, \epsilon)]^{-1}
\end{equation}
 one obtains the 
finite renormalized result (\ref{ex1}),  obtained by cut-off regularization.
Hence   both regularization schemes  yield equivalent renormalized results.

\subsection{The Rank-Two Separable Potential
\label{s2b}}

Next we consider the renormalization of  rank-two separable potential
(\ref{2a}) for $d=3$. In this case 
 the on-shell $K$ matrix is given by \cite{am} 
\begin{equation}\label{107x}
K(k^2)= \frac{u^2(k) (1/\lambda_2- A_{vv})+v^2(k)(1/\lambda_0- A_{uu})
+2u(k)v(k)A_{uv}
  }{(1/\lambda_2- A_{vv})(1/\lambda_0- A_{uu})-A_{uv}^2},
\end{equation}
 where 
 \begin{eqnarray}
 A_{uu}&\equiv&  {\cal P} \int q^2 dq u^2(q)G(q; k^2),\label{a1}\\
   A_{uv}&\equiv&{\cal P}\int q^2 dq u(q)v(q)G(q; k^2),\label{a2}   \\
     A_{vv}&\equiv&{\cal P}\int q^2 dq v^2(q)G(q; k^2).\label{a3}
 \end{eqnarray}
In order to
work out the divergent
terms explicitly 
 we consider  a specific case:
 $u(p)=1,  v(p) =p^2$. 
 This specialization  
 does not correspond to any real loss of generality.
Other choices of $u(p)$ and $v(p)$ can be worked out similarly. 
With this choice the $K$ matrix becomes 
\begin{equation}\label{107}
K(k^2)= \frac{ [1/\lambda_2- I_2(k)]+k^4[1/\lambda_0- I_0(k)]
+2k ^ 2 I_1(k)
  }{[1/\lambda_2- I_2(k)][1/\lambda_0- I_0(k)]-I_1^2(k)}.
\end{equation}
In this case,  divergent integrals  $A_{uu}$,  $A_{uv}$,  and $A_{vv}$  are 
$I_L(k)$ of  (\ref{4b})  with $L=0,  1$ and 2,   respectively. These
 have been treated by dimensional and cut-off regularizations in Sec.
\ref{s2a}. Hence one can use the same regularization procedure as employed
there. However,  in general, 
 $\lambda_0$ and $\lambda_2$ of Eq. (\ref{107})
are to be interpreted as
cut-off ($\Lambda$) dependent bare coupling.  

In this case first we perform  dimensional regularization. From 
Eq. (\ref{32}),  we find that the 
dimensional regularization of integrals $I_L(k)$
for $d=3$  are all zero.
Then, if we use the energy-dependent bare couplings (\ref{bc}), 
from Eq. (\ref{107})
 one obtains the following finite regularized on-shell $K$ matrix
\begin{eqnarray}\label{108}
K_R(k^2)
  ={\Lambda_{00}^{-1}(k^2)+\Lambda_{02}^{-1}(k^2)k^4}, \label{108a}
\end{eqnarray}
where $\Lambda_{0L}^{-1}(k^2)$'s are polynomials in $k^2$ as in Eq.
(\ref{poly}). The renormalized $K$ matrix (\ref{108a}) will have the form 
of a polynomial in $k^2$ at low energies.

Next we employ cut-off regularization in  Eq. (\ref{107}). If we use
Eq.  (\ref{101}),  regularized $K$ matrix (\ref{107}) becomes
\begin{eqnarray}\label{109}
K_R(k^2)&=& \frac{A+Bk^2+Ck^4}{D+Ek^2+Fk^4},\\
&=& \frac{A}{D}+\left(\frac{B}{D}-\frac{AE}{D^2}  \right)k^2+\ldots,
\label{109a}
\end{eqnarray}
where 
\begin{eqnarray}
A &=& \frac{1}{\lambda_2}+\frac{\Lambda^5}{5}, \hskip .5cm B=
-\frac{\Lambda^3}{3} , \hskip .5cm C=\frac{1}{\lambda_0},\\
D&=&\frac{1}{\lambda_0\lambda_2}+\frac{\Lambda^5}{5\lambda_0}
+\frac{\Lambda}{\lambda_2}+\frac {4\Lambda^6}{45}, 
E=\frac{\Lambda^3}{3\lambda_0}-
\frac{\Lambda^4}{3} \hskip .5cm
\makebox{and} \hskip .5cm F=\frac{\Lambda}{\lambda_0}.
\end{eqnarray}
Equation (\ref{109}) represents the exact solution and 
Eq. (\ref{109a}) is its 
low-energy expansion. One performs  renormalization subject to 
conditions $k\to 0$ and $\Lambda \to \infty.$ As there are two couplings 
$\lambda_0$ and $\lambda_2$,  two renormalization conditions can be used in
this case. These two conditions can be used to determine the constant term
and the coefficient of the $k^2$ term in Eq. (\ref{109a}). Thus one introduces
two renormalized parameters via
 \begin{equation}
\lambda_0^R= \lim_{\Lambda\to\infty} \frac{A}{D},  \hskip .5cm \lambda_2^R=
\lim_{\Lambda\to\infty}\left(\frac{B}{D}-\frac{AE}{D^2}  \right)
\label{91} \end{equation}
so that the renormalized $K$ matrix becomes 
\begin{eqnarray}\label{105}
K_R(k^2)
  =\lambda_0^ R+\lambda_2^R k^2+\ldots, \label{105a}
\end{eqnarray}
which is the result of cut-off renormalization at low energies 
up to terms linear in $k^2$. Equations (\ref{91}) define
the bare couplings in the limit $\Lambda \to \infty, $ although it is not
possible to write  closed-form expressions for them in this case.

In this case it is possible to guarantee the equivalence of the $K$
matrices (\ref{108})  and (\ref{109}). For this purpose, it is necessary
to take advantage of the energy-dependence of the bare couplings to ensure 
\begin{eqnarray}
\Lambda_{00}^{-1}(k^2)+
\Lambda_{02}^{-1}(k^2)k^4= \lim_{\Lambda \to \infty}\frac{A+Bk^2+Ck^4}
{D+Ek^2+Fk^4}
\end{eqnarray}
where some care must be taken since bare couplings $\lambda_0$ and $\lambda_2$
may depend on $\Lambda$. One can also enforce equivalence by writing 
$\lambda_0$ and $\lambda_2$ as energy-dependent functions of $\Lambda_{00}$
and $\Lambda_{02}$. A suitable choice of bare couplings will give the 
equivalence of the two results.

\subsection{\label{s2c}A $\delta$-function Potential and its Derivatives}

Now we consider potential (\ref{3})  for $d=3$,  which is the sum of
a $\delta$-function and its second derivatives in configuration space. This
potential appears in the field theoretic reduction of low-energy
nucleon-nucleon potential \cite{coh1,coh2,eft2,wei}. 
After a straightforward calculation,  
 the on-shell $K$ matrix for this potential is given by
\begin{equation}
 K(k^2)= \frac{{2k^2}/{\lambda_2}+{\lambda_1}/{\lambda_2^2}
+I_2(k)-2k^2 I_1(k)+k^4 I_0(k)}
{{1}/{\lambda_ 2^2}-{2I_2(k)}/{\lambda_2}+I_1^2(k)
-{\lambda_1 I_0(k)}/{\lambda_2^2}-I_0(k)I_2(k)},\label{a10}
\end{equation}
where $I_L(k)$'s  are given by  (\ref{4b}). These  integrals have
been treated by dimensional and cut-off regularizations in Sec.
\ref{s2a}. Hence one can use the same regularization procedure as employed
there.   Again,  
 $\lambda_1$ and $\lambda_2$ of Eq. (\ref{a10})
are to be interpreted as
cut-off ($\Lambda$) dependent bare coupling.  

Here, first we perform  dimensional regularization. From Eq.
(\ref{32}),  we find that the dimensional regularization of integrals
$I_L(k)$ for $d=3$  are all zero.  Then, if we use the energy-dependent bare
couplings (\ref{bc}), from Eq. (\ref{107}) one obtains the following finite
regularized on-shell $K$ matrix
\begin{eqnarray}
K_R(k^2)
  ={\Lambda_{1}^{-1}(k^2)+\Lambda_{2}^{-1}(k^2)k^2}, \label{a12}
\end{eqnarray}
where $\Lambda_1^{-1}(k^2)$ and $\Lambda_2^{-1}(k^2)$ 
are polynomials in $k^2$ as in Eq. (\ref{poly}). Hence the 
renormalized $K$ matrix 
(\ref{a12}) has the form 
of a polynomial in $k^2$ at low energies.

Next we employ cut-off regularization in  Eq. (\ref{a10}).
Then the 
quantities $I_0(k)$,  $I_2(k)$ and $I_4(k)$ are,  respectively, 
given by  Eq. (\ref{101}) with $L=0$,  1 and 2,   and we obtain from Eq.
(\ref{a10})
\begin{eqnarray}\label{122}
K_R(k^2)&=& \frac{A+Bk^2}{D+Ek^2},\\
&=& \frac{A}{D}+\left(\frac{B}{D}-\frac{AE}{D^2}  \right)k^2+\ldots,
\label{107a}
\end{eqnarray}
where 
\begin{eqnarray}
A &=& \frac{\lambda_1}{\lambda_2^2}-\frac{\Lambda^5}{5}, \hskip .5cm B=
\frac{\Lambda^3}{3}+\frac{2}{\lambda_2},  \\
D&=&\frac{1}{\lambda_2^2}+\frac{2\Lambda^3}{3\lambda_2}
+\frac{\lambda_1\Lambda}{\lambda_2^2}-\frac {4\Lambda^6}{45}\hskip .5cm
\makebox{and} \hskip .5cm
E=\frac{2\Lambda}{\lambda_2}+
\frac{\Lambda^4}{3}.
\end{eqnarray}
Equation (\ref{122}) represents the exact solution and 
Eq. (\ref{107a}) is its 
low-energy expansion.
Now as in Sec. \ref{s2b} one could introduce two renormalized parameters 
via Eq. (\ref{91}) and obtain a cut-off renormalized 
$K$ matrix at low energies 
as in Eq. (\ref{105a}). The cut-off renormalized 
$K$ matrix  can be 
consistent with the dimensionally renormalized one (\ref{a12}),
if one exploits the flexibility introduced by the energy-dependent 
bare couplings to ensure 
\begin{eqnarray}
\Lambda_{1}^{-1}(k^2)+\Lambda_{2}^{-1}(k^2)k^2
&=& \lim_{\Lambda \to \infty}\frac{A+Bk^2}{D+Ek^2}
\end{eqnarray}
where again 
some care must be taken since bare couplings $\lambda_0$ and $\lambda_2$
may depend on $\Lambda$.

\subsection{\label{s2d} The Tensor Potential}

Finally,  we consider the renormalization of the tensor potential (\ref{40})
in three dimensions.
The $K$ matrix elements in this case satisfy the following set of coupled
equations 
\begin{eqnarray}
K_{LL'}(p, k, k^2) = V_{LL'}(p, k)+ {\cal P}\sum_{L''=0, 2}
\int q^{2} dq V_{LL''}(p, q)
G(q; k^2)K_{L''L'}(q, k, k^2). 
\label{20} \end{eqnarray} 
From Eqs. 
(\ref{40})  and (\ref{20})  it is realized that the $K$ matrix elements
have the following form
\begin{equation}\label{50}
|K_{LL'}(p, q, k^2)| \equiv
\left(\begin{array}{cc} K_{00}(p, q, k^2)   &   K_{02}(p, q, k^2)\\
K_{20}(p, q, k^2)  & K_{22}(p, q, k^2)  \\ \end{array}\right)=
 \left(\begin{array}{cc} \tau_0   &   \tau_1 q^2\\
\tau_1 p^2  & \tau_2 p^2q^2  \\ \end{array}\right),
\end{equation}
where the energy-dependent functions $\tau$'s are defined by
\begin{equation}\label{60}
\left(\begin{array}{cc} \tau_0   &   \tau_1\\
\tau_1  & \tau_2  \\ \end{array}\right)=
\left(\begin{array}{cc} \lambda_0   &   \lambda_1\\
\lambda_1  & \lambda_2  \\ \end{array}\right)
+\left(\begin{array}{cc} \lambda_0   &   \lambda_1\\
\lambda_1  & \lambda_2  \\ \end{array}\right)
\left(\begin{array}{cc} I_0(k)   &   0\\
0  & I_2(k)  \\ \end{array}\right)
\left(\begin{array}{cc} \tau_0   &   \tau_1\\
\tau_1  & \tau_2  \\ \end{array}\right), 
\end{equation}
with $I_0(k)$ and $I_2(k)$  given by Eq. (\ref{4b}). 

Equation (\ref{60})  can be rewritten as 
\begin{equation}\label{80}
\left(\begin{array}{cc} 1-\lambda_0I_0(k)   &   -\lambda_1I_2(k)\\
-\lambda_1I_0(k)  &  1-\lambda_2I_2(k) \\ \end{array}\right)
\left(\begin{array}{cc} \tau_0   &   \tau_1\\
\tau_1  & \tau_2  \\ \end{array}\right)=
\left(\begin{array}{cc} \lambda_0   &   \lambda_1\\
\lambda_1  & \lambda_2  \\ \end{array}\right).
\end{equation}
The following solution of Eq. 
(\ref{80})  can be obtained after straightforward 
algebra
\begin{equation}\label{90}
\left(\begin{array}{cc} \tau_0   &   \tau_1\\
\tau_1  & \tau_2  \\ \end{array}\right)=
\frac{1}{\cal D}
\left(\begin{array}{cc} \lambda_0+(\lambda_1^2-\lambda_0 \lambda_2)I_2(k) 
  &   \lambda_1\\
\lambda_1  & \lambda_2+
(\lambda_1^2-\lambda_0 \lambda_2)I_0(k) \\ \end{array}\right),
\end{equation}
where 
\begin{equation}\label{92}
{\cal D}\equiv [1-\lambda_0I_0(k)][(1-\lambda_2I_2(k)]-\lambda_1^2I_0(k)
I_2(k)
\end{equation}
is the determinant of the first matrix on the left-hand side of Eq. (\ref{80}).
If we use   dimensionally regularized result (\ref{32}) for 
integrals $I_0(k)$ and $I_2(k)$ in Eqs. (\ref{90}) and (\ref{92}),  we obtain
the following dimensionally regularized $\tau$ function:
\begin{equation}\label{93}
\left(\begin{array}{cc} \tau_{R0}   &   \tau_{R1}\\
\tau_{R1}  & \tau_{R2}  \\ \end{array}\right)=
\left(\begin{array}{cc} \Lambda_0^{-1}(k^2)   &   \Lambda_1^{-1}(k^2)\\
\Lambda_1^{-1}(k^2)  & \Lambda_2^{-1}(k^2)  \\ \end{array}\right),
\end{equation} 
where we have employed the usual energy-dependent bare couplings.

Next we employ cut-off regularization to (\ref{90}). If we use 
 cut-off regularized result (\ref{101}) for integrals $I_L(k)$, 
  each of $\tau_0$,  $\tau_1$  and $\tau_2$  will have a form similar to
the right-hand side of Eq. (\ref{109}) and one can make an expansion in $k^2$
as in Eq. (\ref{109a}). Using explicit forms of cut-off regularized  integrals 
$I_L(k)$,  the regularized 
$\tau_0$,  $\tau_1$  and $\tau_2$  can be written as 
\begin{eqnarray}
\tau_{R0} & = &  J[\lambda_0-(\lambda_1^2-\lambda_0\lambda_2)\Lambda^5/5]
+{\cal O}(k^2)+\ldots,\\
\tau_{R1}  & = &  J \lambda_1   +{\cal O}(k^2)+\ldots, \\
\tau_{R2}  &  =  & J[\lambda_2-(\lambda_1^2-\lambda_0\lambda_2)\Lambda]   
+{\cal
O}(k^2)+\ldots,
\end{eqnarray}
where
\begin{equation}
J=[1+\lambda_0\Lambda+\lambda_2\Lambda^5/5-(\lambda_1^2-\lambda_0\lambda_2
)\Lambda^6/5]^{-1}.
\end{equation}
In this case there are three couplings in potential (\ref{40}):
$\lambda_0$,  $\lambda_1$  and $\lambda_2$. These couplings are to be
interpreted as the bare 
couplings for regularization and renormalization. Now if one introduces three
renormalized couplings $ \lambda_{R0}$,  $ \lambda_{R1}$ and $ \lambda_{R2}$
through
\begin{eqnarray}\label{i}
\lim_{\Lambda\to\infty} J[\lambda_0-(\lambda_1^2-\lambda_0
\lambda_2)\Lambda^5/5]&  = & \lambda_{R0},\\ \label{ii}
\lim_{\Lambda\to\infty} J \lambda_1 &  = & \lambda_{R1},\\ \label{iii}
\lim_{\Lambda\to\infty}
J[\lambda_2-(\lambda_1^2-\lambda_0\lambda_2)\Lambda]&  = & \lambda_{R2},
\end{eqnarray}
 one obtains 
the following energy-independent solution valid in the extreme low-energy limit
\begin{equation}\label{97}
\left(\begin{array}{cc} \tau_{R0}   &   \tau_{R1}\\
\tau_{R1}  & \tau_{R2}  \\ \end{array}\right)=
\left(\begin{array}{cc} \lambda_{R0}   &   \lambda_{R1}\\
\lambda_{R1}  & \lambda_{R2}  \\ \end{array}\right).
\end{equation} 
Again it is possible to establish equivalence between results
 (\ref{93}) and (\ref{97}).

\section{\label{s3} Numerical Study }

The renormalization of the
sum of a finite-range and a divergent potential is of considerable interest 
in the context effective-field-theoretic nucleon-nucleon interaction
\cite{coh3,lep,eft2,eft3}.  
Such sum of divergent and
finite-range potentials appear in a field-theoretic description of low-energy
nucleon-nucleon interaction as in (\ref{div}) \cite{eft2,wei}.  It is not
even clear whether such potentials could be successfully renormalized.  We
address this point in the following,  where the finite-range potential is
taken to be a separable one. A general discussion with a complicated
finite-range potential will only add to numerical complication.  We perform
renormalization when a finite-range attractive separable potential, 
$V_f(p, q)=\lambda_1 u(p) u(q)$ with $u(p)= [\alpha^2/(\alpha^2+p^2)]^2, $ is
added to a divergent potential,  $V_d(p, q)$.  
Among divergent potentials we specifically
consider $V_d(p, q)=\lambda_2$ or  $V_d(p, q)=\lambda_2p^2q^2, $   which appears
in Eq. (\ref{div}).

The $K$ matrix for finite-range separable potential $V_f(p, q)$ is
\begin{equation}\label{69}
K(k^2)=\frac {u^2(k)}{1/\lambda_ 1- A_{uu}},
\end{equation}
with $A_{uu}$  given by Eq. 
(\ref{a1}).  Potential $V_f(p, q)$ is taken to model an $S$-wave  spin-triplet
nucleon-nucleon interaction.   We employ nuclear units:
$\hbar^2/2m$ =41.47 MeV fm$^2$,  where $2m$ is the nucleon mass and take
$\alpha^2 = 1 $ fm$^{-2}$. We assume that $V_f(p, q)$
   supports a bound deuteron and produces the
experimental  scattering length  $a\equiv  \pi K(k^2) /2$ = 5.42 fm.
This fixes the value of
coupling to be   $\lambda_1\equiv \lambda_0=-4.9737$ fm.

We  consider the sum of  divergent potential $V_d(p, q)$,  and the separable
potential,   $V_f(p, q)$,  given by Eq. (\ref{3}) with $v(p)=1$ or $v(p)=p^2$ and
$u(p)$ defined above.  The  formal solution for the $K$ matrix with this
potential is given by Eq. (\ref{107x}), where the only divergent integral is
$A_{vv}$.  In order to obtain a regularized $K$ matrix,  we introduce  a bare
coupling in (\ref{107x}) in place of $\lambda_2$ such that
$1/\lambda_2^R\equiv (1/\lambda_2-A_{vv})$ is a finite quantity,  where
$\lambda_2^R$ is the renormalized coupling which is taken to be energy
independent.  Then one obtains the following regularized $K$ matrix
\begin{equation}\label{70}
K_R(k^2)= \frac{u^2(k) (1/\lambda_2^R)+v^2(k)(1/\lambda_1- A_{uu})
+2u(k)v(k)A_{uv}}{(1/\lambda_2^R)(1/\lambda_1- A_{uu})-A_{uv}^2},
\end{equation} 
which is appropriate  for both cut-off and dimensional regularizations.
Of course,  in dimensional regularization,  $A_{vv}=0$  and
$\lambda_2^R=\lambda_2$, if energy-independent bare coupling is used. 
 If we demand that regularized $K$ matrix (\ref{70})
should yield identical phase-shift or scattering length as that produced by
potential
$V_f(p, q), $ we find that $\lambda_2^R=\lambda_2=0$,  which sets the divergent
potential equal to zero.  In order that (\ref{70}) still models the $S$-wave
nucleon-nucleon triplet $K$ matrix in the presence of a non-zero  divergent
potential,  either the strength $\lambda_1$  or the parameter $\alpha$ of the
separable potential should be changed once the divergent  potential is
included. We keep the parameter $\alpha =1$ fm$^{-2}$ unchanged and consider
two possibilities for the coupling $\lambda_1$ of $V_f(p, q)$ in the presence
of the divergent potentials: $
\lambda_1= 1.1\lambda_0= -5.4711$ fm  and $ \lambda_1= 0.9\lambda_0=
-4.4763$ fm. In order that
renormalized $K$ matrix (\ref{70}) leads to the scattering length 5.42 fm,  one
should have $\lambda_2^R = 0.0729$ fm (0.0296 fm)
for $ \lambda_1=1.1\lambda_0$,    and
$\lambda_2^R = -0.0934$ fm ($-$0.0362 fm)
for $ \lambda_1=0.9\lambda_0$ with  the form-factor of divergent 
potential $v(p) =1$ ($p^2$). 
In the case, where the strength of the attractive 
separable interaction 
$V_f(p, q)$ is increased (decreased), 
the renormalized strength of the divergent potential should be 
repulsive (attractive)
in order to reproduce the experimental scattering length 5.42 fm.
This is reflected by the sign of the renormalized strength $\lambda_2^R$ 
of the  divergent potential in both cases. 

The   phase shifts of different schemes are shown in
Fig.  1,  where we plot  phase shifts for potential $V_f(p, q)$ (solid line)
and the two sets of renormalized phase shifts in the presence of the 
divergent potentials  $\lambda_2$ (dashed-dotted line) 
and $\lambda_2p^2q^2$ (dashed line)  for $ \lambda_1=1.1\lambda_0$  and $
\lambda_1=0.9\lambda_0$. The two lines below (above) the solid line 
correspond to $ \lambda_1=1.1\lambda_0$ ($0.9\lambda_0$).

In the presence of ultraviolet divergences,  any renormalization scheme should
lead to physically plausible result  at low energies,  with energies much
lower than the  cut-off. Hence the two versions of renormalized
$K$ matrices (\ref{70})  should be similar to $K$ matrix (\ref{69})
at low energies. From figure 1 we find
that this is indeed the case. An examination of $K$
matrix
(\ref{69}) reveals that it  tends to zero as $k^2 \to \infty$, 
 which is physically
expected from  scattering equation (\ref{2}). However,  $K$ matrix
(\ref{70}) tends to   $\lambda_2^R $ ($\lambda_2^R k^4$) as  $k^2 \to 
\infty$ for 
$v(p) =1$ ($p^2$). Hence at  high
energies the renormalized $K$ matrix has physically unacceptable behavior.
However,  (\ref{70}) yields very reasonable result
at low energies,  which define the domain of validity of renormalization.
In the presence of the divergent potential a new parametrization of the 
interaction is needed to fit the experimental observables. In order to fit 
the phase shifts to higher energies,  specially in the presence of stronger
ultraviolet divergence,  the quantity $1/\lambda_2^R\equiv
(1/\lambda_2-A_{vv})$ has to be taken to be $k^2$ dependent with several 
parameters.

\section{ Summary}

We have renormalized the $K$ matrices obtained with potentials (\ref{1}),
(\ref{2a}),  (\ref{3}) and (\ref{40}) by cut-off and dimensional
regularizations.  
The solution of the dynamical problem in these cases
involves ultraviolet divergences.  For these potentials in three dimensions,
all dimensionally regularized divergent integrals over Green functions are
identically zero.  Both cut-off and dimensional  regularization schemes lead
to equivalent renormalized  results for potentials (\ref{1}),
(\ref{2a}), and (\ref{3}), only 
if general energy-dependent bare couplings are
employed.  For tensor potential (\ref{40}),  both regularizations  in three
dimensions also lead to equivalent $K$ matrices.  
We also performed  renormalization when
the  potential is the sum of a separable and an analytic divergent potential.
For the divergent parts we took one of the following two potentials:
$\lambda$ and $\lambda p^2 q^2$.  The renormalized result is taken to
simulate the $S$-wave spin-triplet nucleon-nucleon potential. We find that in
both cases the low-energy renormalized phase shifts are physically plausible.
The present analytical and numerical studies demonstrate that regularization
and renormalization are efficient tools for treating divergent potentials in
nonrelativistic quantum mechanics.

We thank Dr. Rabin  Banerjee and Dr. Marcelo de M. Leite for  
 informative discussions
and the Conselho Nacional de Desenvolvimento Cient\'{\i}fico e Tecnol\'ogico, 
Funda\c c\~ao de Amparo \`a Pesquisa do Estado de S\~ao Paulo,  
Financiadora de Estudos e Projetos
of Brazil, and the John Simon Guggenheim Memorial Foundation 
 for partial financial support.

\vskip 1.0cm

{\bf Figure Caption:}

1. Phase shifts at different center of mass energies for (a) separable
potential $V_f(p, q)$ (solid line) with coupling $\lambda_0=-4.9737$ fm,  (b)
potential $V_f(p, q)+\lambda_2$ (dashed-dotted line),  and (c) potential
$V_f(p, q)+\lambda_2p^2 q^2$ (dashed line). The phase shifts (b) and (c)
refer to renormalized $K$ matrices. Of the two curves in sets (b) and (c) the
upper (lower) one refers to separable potential coupling $\lambda_1=
0.9\lambda_0$ (1.1$\lambda_0$).

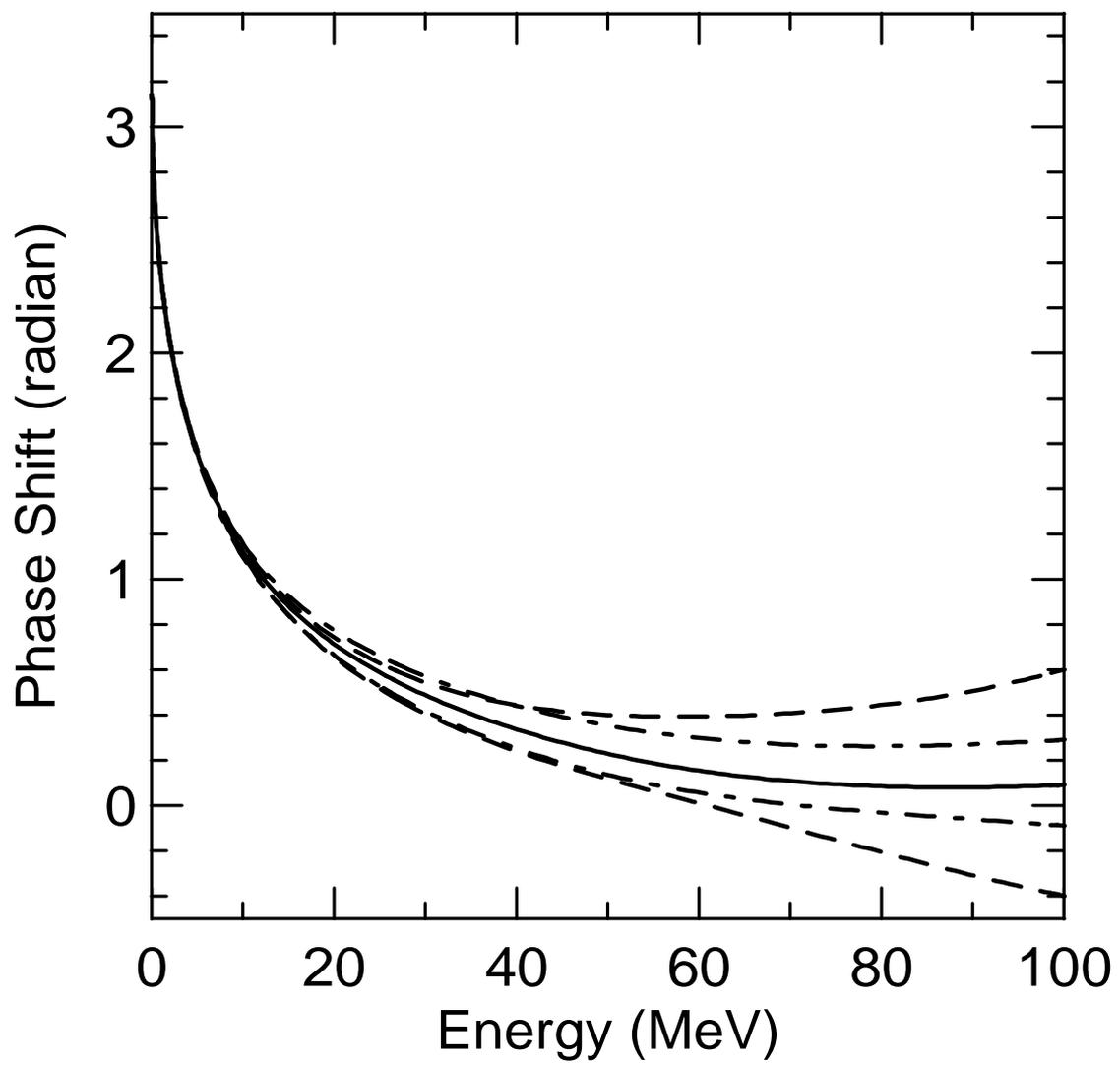
\end{document}